\documentclass[twocolumn,aps,pra,groupedaddress,showpacs]{revtex4}
\usepackage{epsfig,amssymb,amsmath}
\def\comment#1{}

\begin{document}
\title{Gaussian-state quantum-illumination receivers for target detection}
\author{Saikat Guha${}^{1}$, Baris I. Erkmen${}^{2}$}
\affiliation{${}^{1}$Disruptive Information Processing Technologies, BBN Technologies, Cambridge, MA 02138 \\
${}^{2}$California Institute of Technology, Jet Propulsion Laboratory, Pasadena, CA 91109}

\begin{abstract}
The signal half of an entangled twin-beam, generated using spontaneous parametric downconversion, interrogates a region of space that is suspected of containing a target, and has high loss and high (entanglement-breaking) background noise. A joint measurement is performed on the returned light and the idler beam that was retained at the transmitter. An optimal quantum receiver, whose implementation is not yet known, was shown to achieve $6\,$dB gain in the error-probability exponent relative to that achieved with a single coherent-state (classical) laser transmitter and the optimum receiver. We present two structured optical receivers that achieve up to $3\,$dB gain in the error exponent over that attained with the classical sensor. These are to our knowledge the first designs of quantum-optical sensors for target detection, which can be readily implemented in a proof-of-concept experiment, that appreciably outperform the best classical sensor in the low-signal-brightness, high-loss and high-noise operating regime. 
\end{abstract}
\maketitle

A distant region engulfed in bright thermal light, suspected of containing a weakly reflecting target, is interrogated using an optical transmitter. The return light is processed by a receiver to decide whether or not the target is present. Recent work \cite{sacchi2005, lloyd2008, tan2008} has shown that in the above scenario, a ``quantum illumination" (QI) transmitter, i.e., one that generates entangled Gaussian-state light via continuous-wave pumped spontaneous parametric downconversion (SPDC), in conjunction with the optimal quantum receiver, substantially outperforms a coherent-state (un-entangled) transmitter and the corresponding optimum-measurement receiver. This advantage accrues despite the loss of entanglement between the target-return and the idler beams due to the high loss and noise in the intervening medium. This is the first example of an entanglement-based performance gain in a bosonic channel where the initial entanglement does not survive the loss and noise in the system. The SPDC transmitter and optimal receiver combination has been shown to yield up to a factor of $4$ (i.e., $6$ dB) gain in the error-probability exponent over a coherent state transmitter and optimal receiver combination, in a highly lossy and noisy scenario \cite{tan2008}. The optimal receiver for the former source corresponds to the Helstrom minimum probability of error (MPE) measurement \cite{Helstrom1976} under two hypotheses -- $H_0$: target absent, and $H_1$: target present. It can be expressed as a projective measurement onto the positive eigenspace of the difference of the joint target-return and idler density operators under the two hypotheses. However, no known structured optical receiver is yet able to attain the full $6\,$dB predicted performance gain. 

In this paper we present two structured receivers, which, when used in conjunction with the SPDC transmitter, are shown to achieve up to a factor of $2$ error-exponent advantage---i.e., half of the full factor of $4$ predicted by the Helstrom bound---over the optimum-reception classical sensor, in the low signal brightness, high loss and high noise regime. The first receiver uses a low-gain optical parametric amplifier (OPA) and ideal photon counting \cite{guha2009}, whereas the second uses phase-conjugation followed by balanced dual detection. Both receivers attain the same asymptotic error exponent, although the second receiver yields slightly better performance than the first. Both receivers attempt to detect the remnant phase-sensitive cross correlation between the return-idler mode pairs when the target is present~\cite{erkmenshapiro:PCOCT}. Both of our proposed receivers, consisting of separable measurements over $M$ pairs of target-return and idler modes, offer strictly better performance than any classical-state transceiver, and have low-complexity implementations. Apart from the binary-hypothesis target-detection problem considered here, our receivers have been shown to considerably outperform conventional optical sensing and communications applications built on the quantum illumination concept, such as two-way secure communications \cite{Shapiro2009} and standoff one-vs-two-target resolution sensing \cite{Guh2009}. 

Consider $M$ independent signal-idler mode pairs obtained from SPDC, $\{{\hat a}_S^{(k)}, {\hat a}_I^{(k)}\}$; $1 \le k \le M$. Each $T$-sec-long transmission comprises $M = WT \gg 1$ signal-idler mode pairs, where $W$ is the SPDC sourceÕs phase-matching bandwidth. Each mode pair is in an identical entangled two-mode-squeezed state with a Fock-basis representation~\cite{footnote1}
\begin{equation}
|\psi\rangle_{SI} = \sum_{n=0}^{\infty}\sqrt{\frac{N_S^n}{(N_S+1)^{n+1}}}|n\rangle_S|n\rangle_I,
\end{equation}
where $N_S$ is the mean photon number in each signal and idler mode. $|\psi\rangle_{SI}$ is a pure maximally-entangled zero-mean Gaussian state with covariance matrix $V^{SI} = \langle[
\begin{array}{cccc}
{\hat a}_S & {\hat a}_I & {\hat a}_S^{\dagger} & {\hat a}_I^\dagger 
\end{array}
]^{T}[
\begin{array}{cccc}
{\hat a}_S^{\dagger} & {\hat a}_I^\dagger & {\hat a}_S & {\hat a}_I 
\end{array}
]\rangle$ given by
\begin{equation*}
\left [
\begin{smallmatrix}
N_S + 1 & 0 & 0 & \sqrt{{N_S(N_S+1)}} \\
0 & N_S+1 & \sqrt{{N_S(N_S+1)}} & 0 \\
0 & \sqrt{{N_S(N_S+1)}} & N_S & 0 \\
\sqrt{{N_S(N_S+1)}} & 0 & 0 & N_S
\end{smallmatrix}
\right ].
\end{equation*}
Under hypothesis $H_0$ (no target), the target-return mode ${\hat a_R} = {\hat a_B}$, where ${\hat a_B}$ is in a thermal state with mean photon number $N_B \gg 1$. Under hypothesis $H_1$ (target present), ${\hat a_R} = \sqrt{\kappa}{\hat a_S} + \sqrt{1-\kappa}{\hat a_B}$, where $\kappa \ll 1$, and ${\hat a_B}$ is in a thermal state with mean photon number $N_B/(1-\kappa)$, such that the mean noise photon number is equal under both hypotheses. Under $H_1$, each of the $M$ return-idler mode pairs are in a zero-mean Gaussian state, ${\hat \rho}_{RI}^{(1)}$, with the covariance matrix of each mode, $V^{RI} = \langle [
\begin{array}{cccc}
{\hat a}_R & {\hat a}_I & {\hat a}_R^{\dagger} & {\hat a}_I^\dagger 
\end{array}
]^{T}[
\begin{array}{cccc}
{\hat a}_R^{\dagger} & {\hat a}_I^\dagger & {\hat a}_R & {\hat a}_I 
\end{array}
]\rangle$, given by
\begin{equation*}
\left[
\begin{smallmatrix}
{\kappa}N_S + N_B + 1 & 0 & 0 & \sqrt{\kappa{N_S(N_S+1)}} \\
0 & N_S+1 & \sqrt{\kappa{N_S(N_S+1)}} & 0 \\
0 & \sqrt{\kappa{N_S(N_S+1)}} & {\kappa}N_S+N_B & 0 \\
\sqrt{\kappa{N_S(N_S+1)}} & 0 & 0 & N_S
\end{smallmatrix}
\right].
\end{equation*}
Under $H_0$, the joint return-idler state for each of the $M$ mode pairs, ${\hat \rho}_{RI}^{(0)}$, is a product of two zero-mean thermal states $({\hat \rho}_{N_B} \otimes {\hat \rho}_{N_S})$ with mean photon numbers $N_B$ and $N_S$ respectively, viz., $V^{RI} = {\rm{diag}}(N_B+1, N_S+1, N_B, N_S)$. 

\begin{figure}
\begin{center}
\includegraphics[width=8.5cm,angle=0]{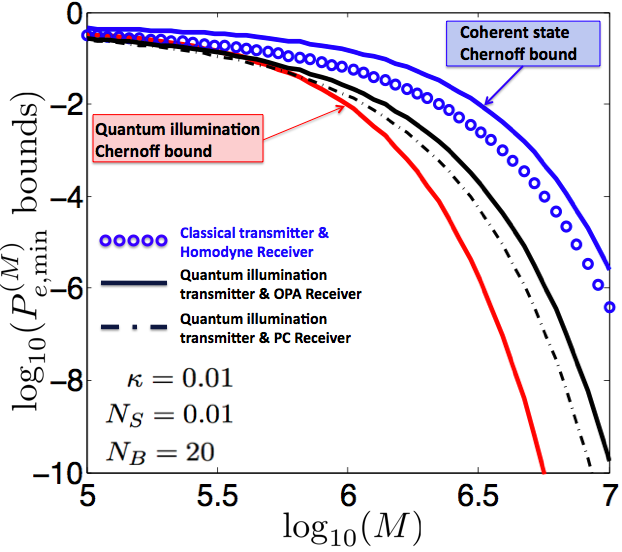}
\end{center}
\caption{(Color online) The figure shows five plots of error-probabilities and bounds thereof as a function of $M$. The two solid curves marked by arrows (from top to bottom respectively) are the coherent-state (blue) and the entangled Gaussian-state (red) transmitters. The third solid curve (black, in between the aforementioned QCB curves) plots the error-probability performance of the OPA receiver, whereas the dash-dotted curve shows the performance of the phase-conjugate receiver. The curve plotted with circles depicts the error-probability performance of the coherent-state transmitter and homodyne detection receiver, which is in fact a lower bound to the performance of an arbitrary classical-state transmitter, including classically correlated signal-idler transmitter states. The parameters used to generate the plots are $N_S=0.01$, $N_B = 20$, and $\kappa = 0.01$.}

\label{fig:bounds}
\end{figure}

The {\em{binary detection problem}} is the MPE discrimination between $H_{0}$ and $H_{1}$ using the optimal measurement on the $M$ return-idler mode pairs, $({\hat \rho}_{RI}^{(m)})^{\otimes M}$ for $m = 0$ or $1$. The minimum probability of error is given by,
$ P_{e,{\min}}^{(M)} = [1 - \sum_n\gamma_n^{(+)}]/2
$, where $\gamma_n^{(+)}$ are the non-negative eigenvalues of $({\hat \rho}_{RI}^{(1)})^{\otimes M} - ({\hat \rho}_{RI}^{(0)})^{\otimes M}$ \cite{Helstrom1976}. The quantum Chernoff bound (QCB), given by $Q_{\rm QCB} \triangleq \min_{0 \le s \le 1}Q_s$ where $Q_s \triangleq {\rm{Tr}}\bigl[({\hat \rho}_{RI}^{(0)})^s({\hat \rho}_{RI}^{(1)})^{1-s} \bigr]$, is an upper bound to $P_{e,{\min}}^{(M)}$ and is asymptotically tight in the exponent of the minimum error probability \cite{audenaert2007}. In particular, we have
\begin{equation}
P_{e,{\min}}^{(M)} \le \frac{1}{2}Q_{\rm QCB}^M \le \frac{1}{2}Q_{0.5}^M,
\label{eq:bounds}
\end{equation}
where the first inequality (QCB) is asymptotically tight as $M \to \infty$~\footnote{A loose lower bound on the error probability is available in~\cite{tan2008}. As the Chernoff bound is asymptotically tight in error exponent to the optimal receiver's performance as $M\rightarrow \infty$, we focus only on the upper bound in this paper.}. The QCB is customarily represented as $P_{e,{\min}}^{(M)} \le e^{-MR_Q}/2$ in terms of an error exponent $R_Q \triangleq -{\ln}(Q_{\rm QCB})$. The second inequality is a looser upper bound known as the Bhattacharyya bound. Symplectic decomposition of Gaussian-state covariance matrices was used to compute the QCB \cite{Pirandola2008, tan2008}, and it was shown that in the high loss, weak transmission and bright background regime, i.e., with $N_S \ll 1$, $\kappa \ll 1$, and $N_B \gg 1$, the entangled transmitter yields a QCB error-exponent $R_Q = {\kappa}N_S/N_B$, which is four times (or $6$ dB) higher than the error-exponent $R_C = {\kappa}N_S/(4N_B)$ for a coherent-state transmitter with a mean photon number $N_{S}$  per mode. In Fig.~\ref{fig:bounds}, we plot the QCB for the entangled and coherent state transmitters, showing a clear advantage of quantum over classical illumination.

When a coherent-state transmitter is used, each received mode ${\hat a_R}$ is in a thermal state with mean photon number $N_B$, and a mean-field $\langle{\hat a_R}\rangle = 0$ or $\sqrt{\kappa{N_S}}$ for hypotheses $H_0$ and $H_1$ respectively. Homodyne detection on each received mode ${\hat a_R^{(k)}}$ yields a variance-$(2N_B+1)/4$ Gaussian-distributed random variable $X_k$ with mean $0$ or $\sqrt{\kappa{N_S}}$ given the hypothesis. The minimum error probability decision rule is to compare $X = X_1 + \ldots + X_M$ against a threshold: $``H_0"$ is declared if $X < (M\sqrt{\kappa{N_S}})/2$ and $``H_1"$ otherwise. The corresponding probability of error is
\begin{equation*}
P_{e,{\rm{hom}}}^{(M)} = \frac{1}{2}{\rm{erfc}}\left(\sqrt{\frac{\kappa{N_S}M}{4N_B+2}}\right) \approx \frac{e^{-MR_{C_{\rm{hom}}}}}{2  \sqrt{ \pi MR_{C_{\rm{hom}}}}},  \nonumber  
\end{equation*}
where $\text{erfc}(x) \triangleq (2/\sqrt{\pi}) \int_{x}^{\infty} e^{-t^{2}} {\rm d}t $, $R_{C_{\rm{hom}}} ={\kappa{N_S}}/({4N_B+2})$ is the error exponent, and the approximation holds for ${\kappa}N_SM/(4N_B+2) \gg 1$. When $N_B \gg 1$, $R_{C_{\rm{hom}}} \approx {\kappa}N_S/4N_B = R_{C}$, so mode-by-mode homodyne detection is asymptotically optimal for the coherent-state transmitter.


\vspace{-0.3cm}
\subsection{The OPA Receiver}

\begin{figure}
\begin{center}
\includegraphics[width=8.5cm,angle=0]{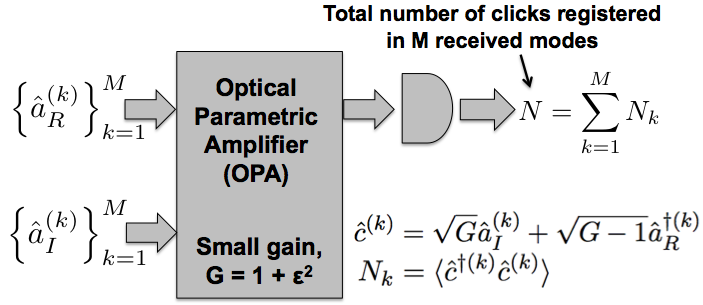}
\end{center}
\caption{In the OPA receiver, the return modes and idler modes are inputs to an optical parametric amplifier (OPA) with gain $G$. The total number of photons, $N$, are counted at one output port. The receiver decides in favor of hypotheses $H_0$ if $N$ is below a threshold $N_{\rm{th}}$, and in favor of $H_{1}$ otherwise.} 
\label{fig:OPArcvr}
\end{figure}

Unlike the coherent-state transmitter, the entangled transmitter results in zero-mean joint return-idler states under both hypotheses. The sole distinguishing factor between the two hypotheses that makes quantum illumination perform superior to the unentangled coherent-state transmitter, are the off-diagonal terms of $V^{RI}$ bearing the remnant phase-sensitive cross correlations of the return-idler mode pairs when the target is present. The OPA receiver uses an optical parametric amplifier to combine the incident return and idler modes ${\hat a}_R^{(k)}$ and ${\hat a}_I^{(k)}$, $1 \le k \le M$, producing output mode-pairs: 
\begin{equation}
{\hat c}^{(k)} = \sqrt{G}{\hat a}_I^{(k)} + \sqrt{G-1}{\hat a}_R^{\dagger{(k)}} 
\end{equation} 
and
\begin{equation}
{\hat d}^{(k)} = \sqrt{G}{\hat a}_R^{(k)} + \sqrt{G-1}{\hat a}_I^{\dagger{(k)}},
\end{equation}
where $G > 1$ is the gain of the OPA (see Fig.~\ref{fig:OPArcvr}). Thus, under both hypotheses ${\hat c}^{(k)}$ is in an independent, identical, zero-mean thermal state,  ${\hat \rho}_c = \sum_{n=0}^{\infty}(N_m^n/(1+N_m)^{1+n})|n\rangle\langle{n}|$, for $m \in \left\{0, 1\right\}$, where the mean photon number is given by $N_{0} \triangleq 
GN_S + (G-1)(1+N_B)$ under $H_{0}$, and  $N_{1} \triangleq  GN_S + (G-1)(1+N_B + {\kappa}N_S) + 2\sqrt{G(G-1)}\sqrt{{\kappa}N_S(N_S+1)}$ under $H_{1}$.

The joint state of the $M$ received modes $\left\{{\hat c}^{(k)}, 1 \le k \le M\right\}$, is the $M$-fold tensor product ${\hat \rho}_c^{\otimes{M}}$, and the $M$-fold product of thermal states is diagonal in the $M$-fold tensor-product of photon-number bases. Therefore, the optimum joint quantum measurement to distinguish between the two hypotheses is to count photons on each output mode ${\hat c}^{(k)}$ and decide between the two hypotheses based on the total photon count $N$ over all $M$ detected modes, using a threshold detector. The probability mass function of $N$ under the two hypotheses is given by
\begin{equation}
P_{N|H_m}(n|H_m) = \left(\begin{array}{c}n+M-1\\n\end{array}\right) \frac{N_m^{n}}{(1+N_m)^{n+M}}, \nonumber
\end{equation}
where $n=0,1,2,\dots$ and $m = 0$ or $1$. The mean and variance of this distribution are $MN_m$ and $M\sigma_m^2$ respectively, where $\sigma_m^2 = N_m(N_m+1)$. The minimum error probability to distinguish between the two distributions $P_{N|H_0}(n|H_0)$ and $P_{N|H_1}(n|H_1)$ using $M$ i.i.d. observations is bounded above by the classical Bhattacharyya bound \cite{guha2009},
\begin{equation}
P_{e,{\rm{OPA}}}^{(M)} \le \frac{1}{2}e^{-MR_B}, 
\end{equation}
where with a small OPA gain $G=1+\epsilon^2$, $\epsilon \ll 1$, the error exponent $R_B$ is given by 
\begin{eqnarray}
R_B &=& \frac{{\epsilon}^2\kappa{N_S(N_S+1)}}{2N_S(N_S+1)+2{\epsilon^2}(1+2N_S)(1+N_S+N_B)} \nonumber \\
&\approx& \kappa{N_S}/2N_B,
\end{eqnarray}
for a choice of $\epsilon^2 = N_S/\sqrt{N_B}$, for $N_S \ll 1$, $\kappa \ll 1$, $N_B \gg 1$ \footnote{A different $\epsilon$ with $N_S/N_B \ll \epsilon^2 \ll 1/N_B$ works as well.}. Therefore by construction, for a weak transmitter operating in a highly lossy and noisy regime, the OPA receiver achieves at least a $3$ dB gain in error exponent over the optimum-receiver classical sensor whose QCB error exponent $R_C = \kappa{N_S}/4N_B$. For $N_S \ll 1$ and $\epsilon \ll 1$, both $N_0$ and $N_1 \ll 1$. Hence, a single-photon detector (as opposed to a full photon-counting measurement) suffices to achieve the performance of the receiver depicted in Fig.~\ref{fig:OPArcvr}. Due to the central limit theorem, for $M \gg 1$, $P_{N|H_m}(n|H_m)$, $m \in \left\{0,1\right\}$ approach Gaussian distributions with mean and variance $MN_m$ and $M\sigma_m^2$ respectively. Hence for $M \gg 1$, 
\begin{equation}
P_{e,{\rm{OPA}}}^{(M)} = \frac{1}{2}{\rm{erfc}}\left(\sqrt{R_{{\rm{OPA}}}M}\right) \approx \frac{e^{-MR_{{\rm{OPA}}}}}{2\sqrt{\pi MR_{\rm{OPA}}}}, \nonumber 
\end{equation}
where an error-exponent $R_{{\rm{OPA}}} = (N_1-N_0)^2/2(\sigma_0+\sigma_1)^2$ can be achieved using a threshold detector that decides in favor of hypothesis $H_0$ if $N < N_{\rm{th}}$, and $H_{1}$ otherwise, where $N_{\rm{th}} \triangleq \lceil{M(\sigma_1N_0+\sigma_0N_1)/(\sigma_0+\sigma_1)}\rceil$. Fig.~\ref{fig:bounds} shows that $P_{e, {\rm OPA}}^{(M)}$ is strictly smaller (by $3$ dB in error-exponent) than $P_{e, {\rm hom}}^{(M)}$ --- the error probability achieved by the coherent state transmitter with a homodyne detection receiver. One can show using convexity arguments that in the high background regime, $P_{e, {\rm hom}}^{(M)}$ is in fact a strict lower bound to the error probability achievable by an arbitrary classical-state transmitter, which includes classically-correlated signal-idler mixed states (i.e., those that admit a Glauber P-representation). 

\vspace{-0.3cm}
\subsection{The Phase-Conjugate Receiver}

\begin{figure}[t]
\begin{center}
\includegraphics[width=8.5cm,angle=0]{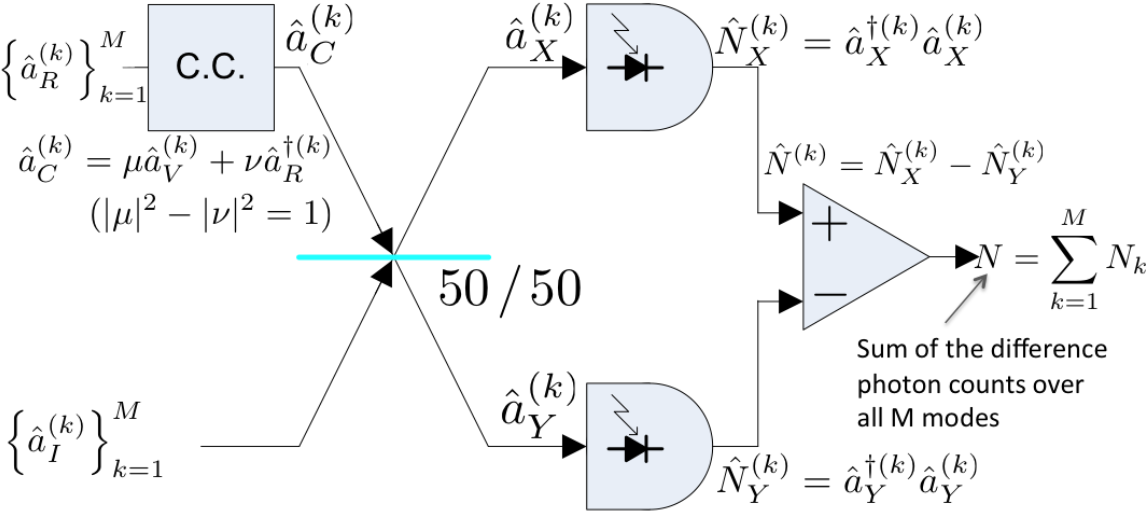}
\end{center}
\caption{In the PC receiver, the phase-conjugated return modes and the idler modes are inputs to a balanced difference detector. If the difference in the total number of clicks, $N$, over all $M$ received modes is less than a threshold $N_{\rm{th}}$, the receiver decides $H_{0}$, and $H_1$ otherwise.} 
\label{fig:PCreceiver}
\end{figure}

The phase-conjugate (PC) receiver is another receiver whose error-probability achieves the same $3$ dB error-exponent gain over the optimal classical transceiver in the asymptotic operating regime $N_S \ll 1$, $\kappa \ll 1$, $N_B \gg 1$, and has slightly better performance than the OPA receiver (see Fig.~\ref{fig:bounds}). As illustrated in Fig.~\ref{fig:PCreceiver}, the receiver phase-conjugates all $M$ return modes ${\hat a}_R^{(k)}$, $1 \le k \le M$ according to 
\begin{equation} 
{\hat a}_C^{(k)} = \sqrt{2}{\hat a}_V^{(k)} + {\hat a}_R^{{\dagger}(k)}\,,
\end{equation}
where ${\hat a}_V^{(k)}$ are vacuum-state operators needed to preserve the commutator. The conjugated return and the retained idler are then detected by a dual, balanced difference detector: the output modes of the 50-50 beam splitter, ${\hat a}_X^{(k)} = ({\hat a}_C^{(k)}+{\hat a}_I^{(k)})/\sqrt{2}$ and ${\hat a}_Y^{(k)} = ({\hat a}_C^{(k)}-{\hat a}_I^{(k)})/\sqrt{2}$, are detected and fed into a unity-gain difference amplifier, such that the final measurement is equivalent to
\begin{equation}
{\hat N}^{(k)} = {\hat N}_X^{(k)} - {\hat N}_Y^{(k)}\,,
\end{equation}
where ${\hat N}_X^{(k)} = {\hat a}_X^{\dagger{(k)}}{\hat a}_X^{(k)}$ and ${\hat N}_Y^{(k)} = {\hat a}_Y^{\dagger{(k)}}{\hat a}_Y^{(k)}$. The final decision is based on the sum of the photon counts $N$ over all $M$ modes.

To simplify the subsequent analysis, let us define ${\bar N}_X \triangleq \langle{{\hat N}_X^{(k)}}\rangle$, ${\bar N}_Y \triangleq \langle{{\hat N}_Y^{(k)}}\rangle$, ${\bar N}_C \triangleq \langle{{\hat a}_C^{\dagger{(k)}}{\hat a}_C^{(k)}}\rangle$, and ${\bar N}_I \triangleq \langle{{\hat a}_I^{\dagger{(k)}}{\hat a}_I^{(k)}}\rangle$. Under hypothesis $H_0$, the modes ${\hat a}_C^{(k)}$ and ${\hat a}_I^{(k)}$ are in product thermal states, whereas under $H_1$ they are in a zero-mean joint Gaussian state with nonzero phase-insensitive cross correlation given by $\langle{{\hat a}_C^{\dagger{(k)}}{\hat a}_I^{(k)}}\rangle = \sqrt{\kappa{N_S}(N_S+1)} = C_q$. Measurement of ${\hat N}^{(k)}$, $1 \le k \le M$, produces a sequence of i.i.d. random variables $N_k$ with mean and variance given by ${N}_0 = 0$ and $\sigma_0^2 = {\bar N}_X({\bar N}_X+1) + {\bar N}_Y({\bar N}_Y+1)$ under hypothesis $H_0$, and ${N}_1 = 2C_q$ and $\sigma_1^2 = {\bar N}_X({\bar N}_X+1) + {\bar N}_Y({\bar N}_Y+1) - ({\bar N}_C - {\bar N}_I)^2/2$ under hypothesis $H_1$. Under hypothesis $H_{0}$ we have ${\bar N}_X = {\bar N}_Y = ({\bar N}_C + {\bar N}_I)/2$, whereas ${\bar N}_X = ({\bar N}_C + {\bar N}_I)/2 + C_q$, and ${\bar N}_Y = ({\bar N}_C + {\bar N}_I)/2 - C_q$ holds for hypothesis $H_1$. Finally, we have ${\bar N}_C = 1 + N_B$ for $H_0$, ${\bar N}_C = 1 + {\kappa}N_S + N_B$ for $H_1$, and because the idler is unaffected under either hypothesis, ${\bar N}_I = N_S$. For large $M$, and hypothesis $H_m$, $m \in \left\{0,1\right\}$, the distribution of $N = \sum_{k=1}^MN_k$ approaches a Gaussian distribution with mean and variance given by $MN_m$ and $M\sigma_m^2$ respectively. Therefore the probability of error
\begin{equation}
P_{e,{\rm{PCR}}}^{(M)} \approx \frac{1}{2}{\rm{erfc}}\left(\sqrt{R_{{\rm{PCR}}}M}\right) \approx \frac{e^{-MR_{{\rm{PCR}}}}}{2\sqrt{\pi MR_{\rm{PCR}}}}, \nonumber 
\end{equation}
where an error-exponent $R_{{\rm{PCR}}} = (N_1-N_0)^2/2(\sigma_0+\sigma_1)^2$ can be achieved using a threshold detector that decides in favor of hypotheses $H_0$ when $N < N_{\rm{th}}$ and in favor of $H_{1}$ otherwise, where $N_{\rm{th}} = \lceil{M(\sigma_1N_0+\sigma_0N_1)/(\sigma_0+\sigma_1)}\rceil$. The corresponding error-exponent is given by 
\begin{eqnarray}
R_{PCR} &=& \frac{\kappa{N_S}(N_S+1)}{2N_B + 4N_SN_B+6N_S+4\kappa{N_S^2}+3{\kappa}N_S+2} \nonumber\\
&\approx& \kappa{N_S}/2N_B,
\end{eqnarray}
where $N_S \ll 1$, $\kappa \ll 1$, $N_B \gg 1$. 

The PC receiver achieves the same $3$ dB error-exponent gain as the OPA receiver over the optimum-reception classical transceiver, though the performance of the former is slightly better in absolute terms (see Fig.~\ref{fig:bounds}). One reason for this is that balanced dual-detection cancels the common-mode excess noise in $\hat{a}_{X}$ and $\hat{a}_{Y}$, which is reflected by the negative term $({\bar N}_C - {\bar N}_I)^2/2$ in the variance of $N_k$ under hypothesis $H_1$. On the other hand, the OPA receiver operates at very low gain, thus requires much less pump power than unity-gain phase-conjugation. 

In summary, we have proposed two receiver structures, both viable for low-complexity proof-of-concept experimental demonstrations using off-the-shelf optical components, which in conjunction with the SPDC entangled-state source, could substantially outperform classical transceivers for various entangled-state optical sensing applications, such as standoff target detection, one-vs-two-target resolution sensing \cite{Guh2009} and two-way secure communications \cite{Shapiro2009}.

\vspace{4mm}
The authors thank J. H. Shapiro for making the observation about generalizing the coherent-state performance bound to arbitrary signal-idler classically-correlated transmitters. The authors also thank F. Wong, S. Lloyd, and Z. Dutton for valuable discussions. SG thanks the DARPA Quantum Sensors Program and BBN Technologies. BIE's contribution to the research described in this paper was carried out at the Jet Propulsion Laboratory, California Institute of Technology, under a contract with the National Aeronautics and Space Administration.


\begin{thebibliography}{10}
\bibitem{sacchi2005} M. F. Sacchi, Phys. Rev. A {\bf{71}}, 062340 (2005); M. F. Sacchi, Phys. Rev. A {\bf{72}}, 014305 (2005).
\bibitem{lloyd2008} S. Lloyd, Science {\bf{321}}, 1463 (2008).
\bibitem{tan2008} S.-H. Tan, B. I. Erkmen, V. Giovannetti, S. Guha, S. Lloyd, L. Maccone, S. Pirandola, and J. H. Shapiro, Phys. Rev. Lett. {\bf{101}}, 253601 (2008).
\bibitem{Helstrom1976} C. W. Helstrom, Mathematics in Science and Engineering {\bf{123}}, {\em Quantum 
Detection and Estimation Theory}, Academic, New York, (1976). 
\bibitem{guha2009} S. Guha, Proceedings of the IEEE International Symposium on Information Theory (ISIT), Seoul, Korea, Jun 28 - Jul 3, 2009. Paper available online: {\textit{www.ieee.org/ieeexplore}}.
\bibitem{erkmenshapiro:PCOCT} B. I. Erkmen and J. H. Shapiro, Phys. Rev. A \textbf{74}, 041601(R) (2006).
\bibitem{Shapiro2009} J. H. Shapiro, Phys. Rev. A  {{\bf 80}}, 022320 (2009).
\bibitem{Guh2009} J. H. Shapiro and S. Guha, NIST Single Photon Workshop (SPW), Nov. 3-6, 2009. Abstract available online: {\textit{photon.jqi.umd.edu/spw2009/}}. 
\bibitem{footnote1} {We drop the superscripts $(k)$ whenever convenient. As each signal-idler mode pair is identical and undergoes identical channel transformation, there is no ambiguity.}
\bibitem{audenaert2007} K. M. R. Audenaert, {\em{et. al.}}, Phys. Rev. Lett. {\bf{98}}, 160501 (2007).
\bibitem{Pirandola2008} S. Pirandola and S. Lloyd, Phys. Rev. A {\bf{78}}, 012331 (2008). 
\end{thebibliography}
\end{document}